\pdfoutput=1

\documentclass[aps,prb,reprint,superscriptaddress]{revtex4-2}
\usepackage{graphicx}
\usepackage{dcolumn}
\usepackage{bm}
\usepackage{epstopdf}
\usepackage{color}
\usepackage{hyperref}
\draft 
\begin{document} 
\title{Dark-state impact on the exciton recombination of WS$_2$ monolayers as revealed by multi-time-scale pump-and-probe spectroscopy} 
%
%
%
\author{Takashi~Kuroda}
\email[Author to whom correspondence should be addressed: ]
{kuroda.takashi@nims.go.jp}
\affiliation{National Institute for Materials Science, 1-1 Namiki, Tsukuba 305-0044, Japan}
\author{Yusuke~Hoshi}
\affiliation{Tokyo City University, 8-15-1 Todoroki, Setagaya-ku, Tokyo 158-0082, Japan}
\affiliation{Institute of Industrial Science, University of Tokyo, 4-6-1 Komaba, Maguro-ku, Tokyo 153-8505, Japan}
\author{Satoru~Masubuchi}
\affiliation{Institute of Industrial Science, University of Tokyo, 4-6-1 Komaba, Maguro-ku, Tokyo 153-8505, Japan}
\author{Mitsuhiro~Okada}
\author{Ryo~Kitaura}
\affiliation{Department of Chemistry, Nagoya University, Nagoya 464-8602, Japan}
\author{Kenji~Watanabe}
\affiliation{National Institute for Materials Science, 1-1 Namiki, Tsukuba 305-0044, Japan}
\author{Takashi~Taniguchi}
\affiliation{National Institute for Materials Science, 1-1 Namiki, Tsukuba 305-0044, Japan}
\affiliation{Institute of Industrial Science, University of Tokyo, 4-6-1 Komaba, Maguro-ku, Tokyo 153-8505, Japan}
\author{Tomoki~Machida}
\affiliation{Institute of Industrial Science, University of Tokyo, 4-6-1 Komaba, Maguro-ku, Tokyo 153-8505, Japan}

\date{\today}

\begin{abstract}
The luminescence yield of transition metal dichalcogenide monolayers frequently suffers from the formation of long-lived dark states, which include excitons with intervalley charge carriers, spin-forbidden transitions, and a large center-of-mass momentum located outside the light cone of dispersion relations. Efficient relaxation from bright exciton states to dark states suppresses the quantum yield of photon emission. In addition, the radiative recombination of excitons is heavily influenced by Auger-type exciton-exciton scattering, which yields another nonradiative relaxation channel at room temperature. Here, we show that Auger-type scattering is promoted not only between (bright) excitons but also between excitons and long-lived dark states. We studied the luminescence dynamics of monolayer WS$_2$ capped with hexagonal BN over broad time ranges of picoseconds to milliseconds using carefully designed pump-and-probe techniques. We observed that luminescence quenching associated with Auger-type scattering occurs on 1-100 $\mu$s time scales, which thus correspond to the lifetimes of the relevant dark states. The broad distribution of the measured lifetimes implies the impact of various types of long-lived states on the exciton annihilation process. 
\end{abstract}
\maketitle 
\noindent
\textcolor{blue}{%
This is the version of the article accepted for publication in \textit{Physical Review B}. The final published version will be available from the journal's site. %
}%
\section{Introduction }%
The high yield light emission associated with exciton recombination makes transition metal dichalcogenide (TMD) monolayers as unique platforms on which to demonstrate various exciting physics phenomena %
\cite{QHWang_NatNano12,Butler_ACSNano13,GeimGrigorieva_Nat2013,Xia_NatPhoton14,Xu_NatPhys14,Yu_NSR15,CastellanosGomez_2016,Mak_NatPhoton2016,Wang_RMP2018}. %
Mechanisms behind the strong light emission include 1) strong interband transitions since both conduction and valence electron states consist of strongly localized metal \textit{d}~orbitals, and the absorption coefficients of TMD commonly exceed 10\% per monolayer depending on wavelength \cite{Mak_PRL10,Splendiani_NanoLett10}, and 2) large exciton binding energies of the order of 0.5~eV \cite{Cheiwchanchamnangij_PRB12,Ramasubramaniam_PRB12,Qiu_PRL13,Chernikov_PRL14,He_PRL14,Ugeda2014,Wang_PRL15}, which ensures the stable formation of excitons even at room temperature. The observation of ultrafast spontaneous emissions with decay times of a few ps confirms the enhanced exciton oscillator strengths %
\cite{Korn_APL11,Wang_PRB14,Wang_APL15,Moody_NatComm2015,Palummo_NanoLett2015,Poellmann_NatNano2015,Jakubczyk_NanoLett2016,Robert_PRB16}. %
However, it is also commonly accepted that standard monolayer TMD samples exhibit a quantum yield lower than 1\%, possibly limited by the presence of crystalline defects \cite{Wang-NL2014}, relaxation to various dark states, which include intervalley excitons, spin-forbidden triplet excitons, and excitons with non-zero center-of-mass momentum, as well as Auger-type scattering that leads to nonradiative exciton annihilation \cite{Kumar_PRB14,Mouri_PRB14,Sun_NL2014,Yuan_Nanoscale15}. Of these phenomena, Auger scattering, i.e., nonradiative energy transfer between excitons, is regarded as a dominant reason for the room-temperature quantum yield being limited, as the effect appears even for moderate densities lower than $10^{10}$~cm$^{-2}$ \,\cite{Amani_Science2015,Yu_PRB16}. The origin of the efficient Auger process is not fully understood, but it constitutes a key issue as regards developing practical light emitting devices working at room temperature. 

Recently we observed that Auger-type exciton annihilation is greatly suppressed by the encapsulation of tungsten disulfide (WS$_2$) monolayers with hexagonal boron nitride (hBN), which is a 
useful substrate for two-dimensional (2D) materials thanks to its reduced surface roughness and low background-carrier densities 
\cite{Hoshi_PRB17}. The improved surface homogeneity of TMD/hBN leads to the formation of delocalized excitons, which are homogeneously dispersed over the 2D plane. In contrast, TMDs supported on a common class of substrates (such as SiO$_2$) tend to have dense localization centers, which increase the local exciton density and enhance the probability of contact-type exciton-exciton scattering. 

In this work we focus on the mechanism of efficient exciton annihilation in TMD monolayers at room temperature. We extended the measurement time scale up to milliseconds by using newly designed luminescence-based pump-and-probe techniques, which made it possible to access hidden slow dynamics over broad time scales. Hence, we revealed efficient Auger-type scattering between excitons and other long-lived ``dark'' states whose lifetimes are much longer than those of bright excitons. The measured lifetimes were distributed in 1-100~$\mu$s time scales, implying that various types of long-lived states contribute to exciton annihilations, which thus led to the emergence of the previously reported low-injection Auger scattering \cite{Amani_Science2015,Yu_PRB16,Hoshi_PRB17}. 

This paper is organized as follows. Section~\ref{sec_sample} describes sample preparation. In Sec.~\ref{sec_psStudy}, we measure the picosecond time-scale response after short-pulsed excitation and confirm that long-lived dark states evidently cause Auger scattering. In Sec.~\ref{sec_qcwStudy}, we study the microsecond time-scale response using a fast modulated quasi-cw source. Then, we directly monitor the accumulation and relaxation process of the long-lived states and quantify their lifetimes. In Sec.~\ref{sec_discussion}, we discuss the identification of the measured long-lived states. 

\section{\label{sec_sample}Samples and setups}%
\begin{figure}
\includegraphics[width=8.2cm]{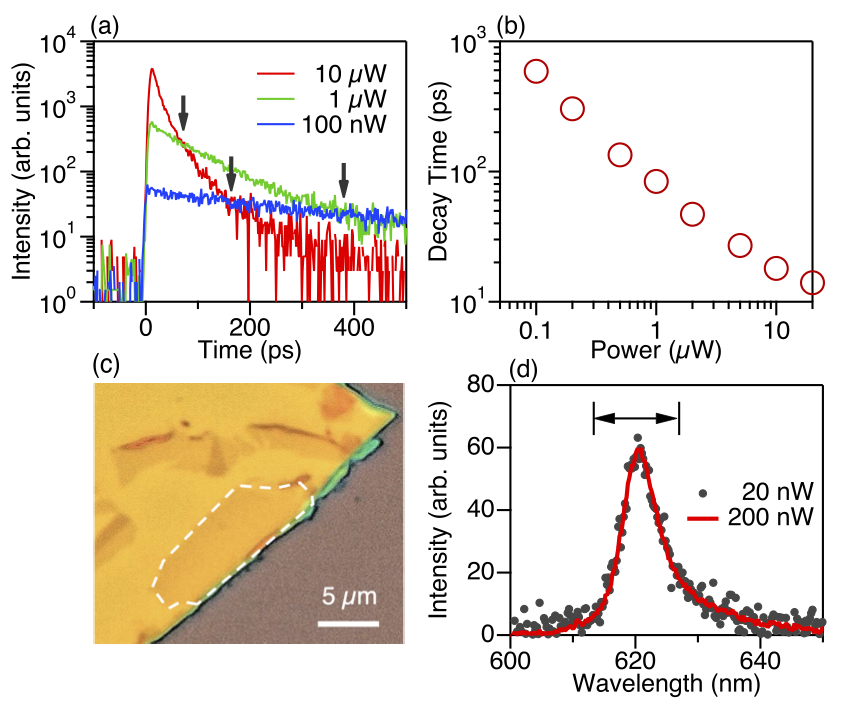}
\caption{\label{fig_decays} Picosecond luminescence dynamics: (a) Time-resolved luminescence signals of monolayer WS$_2$ capped with hBN, excited by 2-ps pulses with 76-MHz repetition rates for various excitation powers. The measurements were performed at room temperature. (b) Decay time dependence on excitation power. (c) Microscope image of our sample. The white broken line highlights the monolayer region. (d) Luminescence spectra. The two signals measured at 20 and 200~nW are normalized to their maxima. The arrow indicates the spectral integration window we used for our time-resolved analysis. }
\end{figure}

We used a CVD grown WS$_2$ monolayer transferred onto a SiO$_2$ substrate through a pick-up transfer process \cite{Onodera_2020}. It revealed significant luminescence quenching at moderate excitation \cite{Hoshi_PRB17}. %
The WS$_2$ monolayer was capped with hBN film to avoid optical damage during the experiment. The sample preparation is detailed in Refs.~\onlinecite{Hoshi_PRB17,Hoshi2018}. (A microscope image of our sample can be seen in Fig.~\ref{fig_decays}(c).) 
In the experiment, we used a home-built confocal setup to collect luminescence signals within a diameter of around 1~$\mu$m on the monolayer flake region. The excitation and detection scheme will be described in later sections. All the experiments were performed at room temperature. 

\section{\label{sec_psStudy}Study with high-repetition short-pulsed excitation source}
In our first set of experiments, we excited the sample with ps light pulses with a duration of 2~ps, a wavelength of 550~nm, and a repetition rate of 76~MHz, which were generated by an optical parametric oscillator. Then, the luminescence signals were analyzed temporally using a synchronously scanning streak camera with a maximum resolution of 2~ps. 

\subsection{Picosecond luminescence decay dynamics}
Figure~\ref{fig_decays}(a) shows the decay curves of the neutral exciton line observed at 
620~nm for different excitation powers. 
(The luminescence spectra are shown in Fig.~\ref{fig_decays}(d).) For the lowest excitation at 100~nW, the decay signal showed a single exponent decay with a time constant of 590~ps (blue line). When the excitation power was increased to 1~$\mu$W, the signal decayed faster with a time constant of 85~ps (green line). A further increase in the excitation power to 10~$\mu$W resulted in a short time constant of 18~ps. (red line. The decay curve at 10~$\mu$W deviates from a single exponent curve. Thus, we fitted the data using a double exponent function and evaluated the decay time as the weighted average of extracted time constants.) 
The decay time dependence on excitation power is shown in Fig.~\ref{fig_decays}(b). 
Such a reduction in lifetime with excitation power is a signature feature of Auger-type exciton annihilation, where excitons are recombined nonradiatively via contact scattering with other excitons \cite{Kumar_PRB14,Mouri_PRB14,Sun_NL2014,Yuan_Nanoscale15,Amani_Science2015,Yu_PRB16}. It is noteworthy that the phenomenon appears at excitation power as low as 100~nW, from which we estimate a photoinjection density of the order of $10^9$~cm$^{-2}$, much smaller than commonly encountered 2D exciton densities that induce nonlinear scattering events in II-VI and III-V quantum well systems \cite{Klingshirn}. 

Note that the luminescence intensity at time origin (0~ps in this figure) is nearly proportional to the excitation power, while the decay constant changes monotonically. Hence, higher excitation curves fall below the lower excitation curves at particular delay times depicted by the vertical arrows. The observed ``signal crossing'' behavior suggests that the luminescence intensity does not simply follow the transient population of excitons generated by each pulse. Instead, the luminescence decay is influenced by previous pulses that accumulate the population of long-lived states whose lifetimes are longer than the exciton lifetime. 

\begin{figure}
\includegraphics[width=7.8cm]{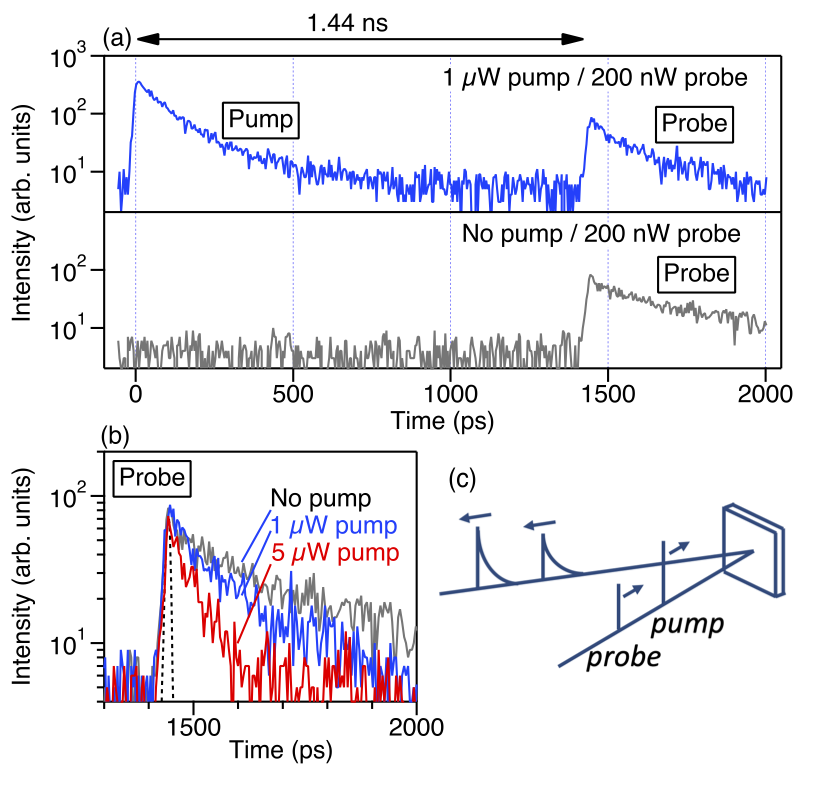}
\caption{\label{fig_2pulse} Luminescence-based pump-and-probe measurement: (a) The top panel shows a luminescence trace excited with strong pump pulses (1~$\mu$W) followed by weak probe pulses (200~nW). The pulse separation is set at 1.44~ns, so that the pump-induced luminescence has entirely finished before the probe pulses are injected. The bottom panel shows a luminescence trace excited with probe pulses alone. (b) Pump-power dependence of probe-induced signals. The broken line shows an instrumental response curve, which we measured with ps excitation pulses. The full-width-at-half-maximum is estimated to be 11~ps, which gives the time resolution of this setup. (c) Measurement scheme. }
\end{figure}

\subsection{\label{subsec_PP}Pump-and-probe measurement while monitoring luminescence decay curves}
To confirm the above hypothesis, we carry out a pump-and-probe-type measurement, where we divide excitation pulses into two pulses and study the impact of preceding ``pump'' pulses on the luminescence induced by delayed ``probe'' pulses. (See the measurement scheme in Fig.~\ref{fig_2pulse}(c).) Figure~\ref{fig_2pulse}(a) shows a time trace of luminescence signals excited by pulse pairs. The pulse separation is set at 1.44~ns, which is sufficiently longer than the measured luminescence decay time ($\leq600$~ps). Nevertheless, the luminescence signals generated by both the pump and probe pulses are simultaneously analyzed in the same streak camera curve that covers the measurement window of $\sim2.1$~ns. Figure~\ref{fig_2pulse}(b) is a summary of the pump power dependence of the decay curves of probe generated signals. The decay time decreases monotonically with increasing pump power, while the peak intensity at time origin remains unchanged, in contrast to the previous observations using single excitation pulses shown in Fig.~\ref{fig_decays}. Hence, the luminescence decay is crucially influenced by the presence of long-lived ``dark'' states, which have been injected by much earlier pump pulses. 

Tuning the time separation between the pump and probe pulses could allow us to detect dark-state relaxation. However, the probe luminescence was not significantly dependent on 
the pulse separation up to 11~ns, which was limited by the repetition rate of our light source. The measurement result is briefly described in 
Appendix. Thus, we conclude that the relevant dark states have lifetimes longer than the 10~ns time scales. 

\section{\label{sec_qcwStudy}Study with fast modulated quasi-CW light source}
To examine such slow dark-state dynamics, we introduce an alternative scheme that utilizes quasi-cw light as an excitation source. Output from a cw laser with a wavelength of 532~nm was temporally modulated using a high-speed TeO$_2$ acousto-optic modulator (AOM, center frequency of 200~MHz, Panasonic, EFLM200) to form square-wave pulses with sharp rising/falling edges (rise time shorter than 10~ns). Then, we measured the time evolution of the ``steady-state'' luminescence intensity using a single-photon detector (PerkinElmer, SPCM-AQR) and a fast multichannel scaler with a minimum time bin of 5~ns (Becker \& Hickl, MSA-300). 

\subsection{Microsecond luminescence quenching under square-wave excitation}

\begin{figure}
\includegraphics[width=6cm]{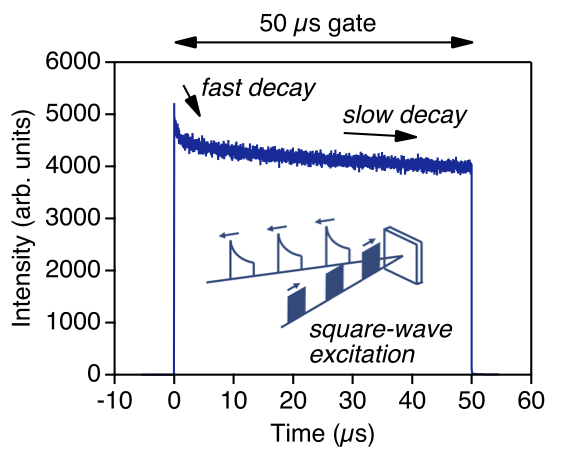}
\caption{\label{fig_cwIntensity} Microsecond luminescence dynamics: Time evolution of luminescence signals under square-wave excitation with a duration of 50~$\mu$s and a duty ratio of 10\%. The square pulses are formed of cw light with a (peak) power of 500~nW and a wavelength of 532~nm. The luminescence evolution is analyzed with a time bin of 10~ns. The measurement scheme is shown in the inset.}
\end{figure}

Figure~\ref{fig_cwIntensity} shows the time evolution of luminescence signals under square-wave excitation with a gate width of 50~$\mu$s and a duty ratio of 10\% (excitation period of 0.5~ms). It reveals significant luminescence quenching with time, along with an accumulation of the dark state population, i.e., the number of excited dark states increases with time after the start of photoinjection, the probability of Auger scattering increases, and the luminescence intensity decreases. 

The signal trace in Fig.~\ref{fig_cwIntensity} consists of two distinct components with different characteristic times. The fast decaying component appears as a spike signal just after the photoinjection began. The slower decaying component, which decreased gradually with time, had not reached its saturation value even at the time the photoinjection was stopped. These findings imply contributions by different long-lived states, which have lifetimes of 
the orders of 1~$\mu$s and $\gg$50~$\mu$s, to exciton luminescence quenching. 

\subsection{Pump-and-probe measurement while monitoring the luminescence recovery}%

\begin{figure}
\includegraphics[width=7cm]{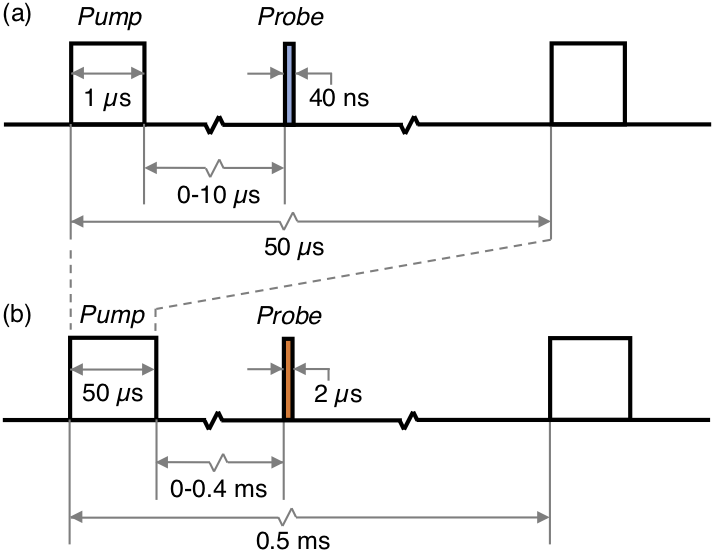}
\caption{\label{fig_timing} 
Timing diagram of pump and probe gate operations for measuring (a) $\mu$s time-scale response and (b) sub-ms time-scale response. %
}
\end{figure}

Dark-state relaxation can be directly explored by monitoring the recovery of the luminescence yield, 
which is accompanied by a progressive decrease in the dark-state density. To accomplish this task we operated AOM to form broad pump gates followed by much narrower probe gates and measured the probe-induced luminescence intensity as a function of delay time. The relative timing of the pump and probe gates was systematically controlled using a digital pulse generator (Quantum Composers, 9200 Plus). The timing diagram of the gate sequence is shown in Fig.~\ref{fig_timing}. In the following we observe the time trace of luminescence signals excited with a controlled gate sequence and analyze the probe-induced luminescence intensity, which becomes higher as the probe pulses are further delayed. 

\subsubsection*{Microsecond time-scale response}

\begin{figure}
\includegraphics[width=6cm]{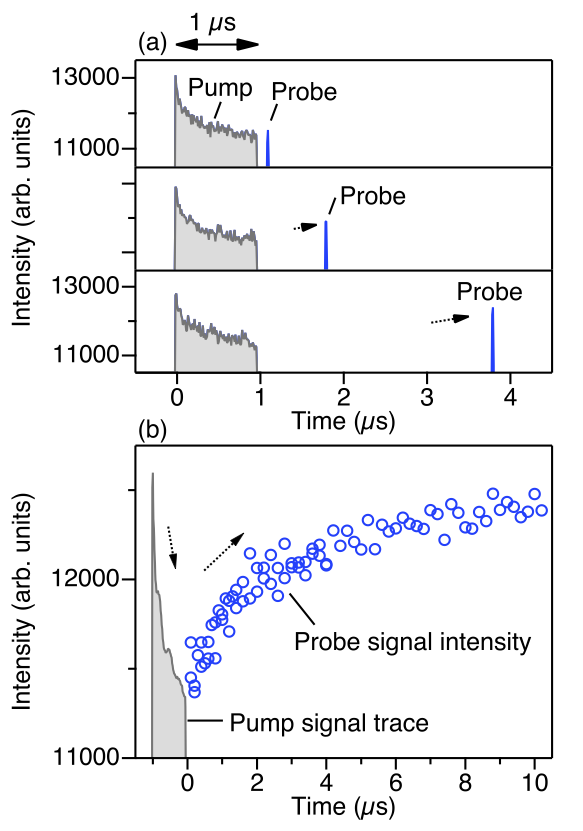}
\caption{\label{fig_PP1microSec} Microsecond pump and probe measurement. (a) Time trace of luminescence signals excited by 1-$\mu$s square pump pulses followed by 40-ns probe pulses for different delay times. See Fig.~\ref{fig_timing}(a) for the gate sequence. (b) Probe-induced luminescence as a function of delay time revealing an intensity recovery with a time constant of 4~$(\pm 1)$~$\mu$s (blue circles). A pump-induced signal trace is also shown with a gray line. %
}
\end{figure}

First, we focus on the rapid dynamics that led to the spike signal in the 50-$\mu$s gate curve (Fig.~\ref{fig_cwIntensity}), which suggested the presence of dark states with a lifetime of the order of 1~$\mu$s. Thus, we set relatively narrow pump and probe gates with durations of 1~$\mu$s and 40~ns, respectively (repetition period of 50~$\mu$s). See the gate timing in Fig.~\ref{fig_timing}(a). %
Figure~\ref{fig_PP1microSec}(a) shows the intensity trace of luminescence signals excited with the controlled gate sequence. The intensity trace 
clearly reveals the light emission at the pump gate (gray curve) and that at the probe gate (blue curve). The pump induced signal exhibits a significant temporal decay, which reproduces the spike signature in Fig.~\ref{fig_cwIntensity}. The probe induced signal increases with delay time towards the initial equilibrium intensity observed at the moment the pump gates were opened, as expected. 

Figure~\ref{fig_PP1microSec}(b) is a summary of probe signal intensities as a function of delay time (blue circles). 
The time trace of the pump-induced signals is also plotted by the gray line. The downward and upward arrows seen above the data curves indicate the progressive change in the luminescence yield, accompanied by the accumulation and relaxation of long-lived dark states, respectively. 
The probe luminescence intensity is recovered with a time constant of $4 \,(\pm 1)$~$\mu$s, which thus gives the lifetime of the relevant dark states that yielded the fast decaying component in the intensity curve in Fig.~\ref{fig_cwIntensity}. %

\subsubsection*{Sub-millisecond time-scale response}%

\begin{figure}
\includegraphics[width=6cm]{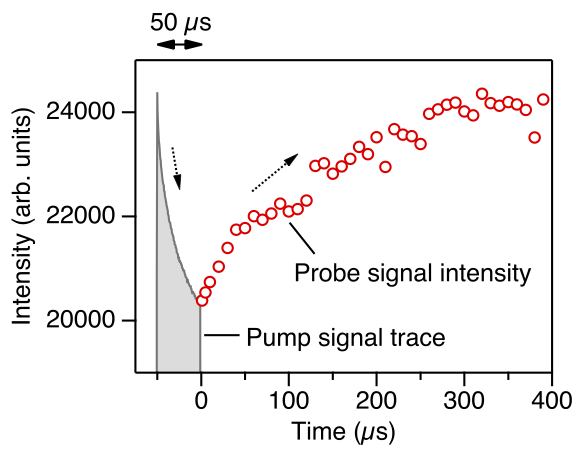}
\caption{\label{fig_PP100microSec} Sub-millisecond dynamics. The result of pump and probe measurements adopting excitation sequences with 50-$\mu$s pump gates followed by 2-$\mu$s probe gates (0.5-ms repetition period). See Fig.~\ref{fig_timing}(b) for the excitation scheme. The trace of the pump-induced luminescence is indicated by a gray line, and the intensity of the probe-induced luminescence is plotted with red open circles. }
\end{figure}

Next, we focus on the slower dynamics that led to a gradual decrease in the intensity curve in Fig.~\ref{fig_cwIntensity}. The pump and probe gates are extended and given durations of 50 and 2~$\mu$s, respectively (repetition period of $0.5$~ms). In this case, the dark state with a 4~$\mu$s lifetime is entirely saturated when the pump gates are opened. See the gate timing in Fig.~\ref{fig_timing}(b). Figure~\ref{fig_PP100microSec} shows the measurement results we obtained when we adopted these excitation sequences, and they reveal the clear recovery of the probe luminescence intensity on a sub-ms scale. The time constant is roughly estimated to be $250 \, (\pm 50)$~$\mu$s, which is thus the lifetime of the relevant states that caused the slow luminescence quenching. 

It should be mentioned that of the two distinct long-lived states with lifetimes of 4 and 250~$\mu$s, the longer lifetime state likely dominates the source of the luminescence quenching under stationary (and moderate) injections. This is because the steady-state (saturated) population of a long-lived state is proportional to its lifetime, and the longer lifetime state is more populated than the other. Hence, exciton annihilations at room temperature are predominantly driven by scattering with long-lived states that have lifetimes on a sub-ms time scale.

%
\section{\label{sec_discussion}Discussion: Identification of measured long-lived states}
In the following we discuss three issues with respect to identifying the observed long-lived states. 
First, we address the role of \textit{exciton center-of-mass motions.} Our study focuses on room-temperature dynamics. Hence, excitons are thermally distributed along different momentum states, and only the small portions located inside the light cone of dispersion relations are allowed to emit light \cite{Robert_PRB16}. The other states with finite in-plane momenta propagate along the 2D layer, and they cannot emit light into free space. Since momentum scattering occurs on time scales shorter than a few picoseconds, as the relevant process does not require spin flip \cite{Selig2016}, we can assume that both bright and dark excitons are fully thermalized regarding center-of-mass motions during our observations. In this case, thermalized excitons are expected to decay with an average lifetime, which is much longer than the intrinsic lifetime determined by the oscillator strength. Our observation of relatively long luminescence lifetimes (590~ps for a weak-excitation limit) compared with reported low-temperature values (a few picoseconds) arises due to the thermal equilibrium condition. %

Second, we consider \textit{spin-forbidden states.} The lowest energy excitons in tungsten dichalcogenide monolayers are known to be optically inactive as they comprise spin triplet configurations \cite{Glazov_PRB14}. A study of high-field magneto-photoluminescence in WS$_2$ monolayers confirmed a dark-bright splitting of 47~meV \cite{Molas_2017}, which exceeds the room-temperature thermal energy. Hence, the majority of excitons stay in spin-forbidden dark states even at room temperature. In view of this fact, we attribute the observed long-lived state that has a lifetime of 4~$\mu$s (Fig.~\ref{fig_PP1microSec}) to spin-forbidden dark excitons. Note that recent observations of intervalley biexcitons, which consist of bright and dark excitons and thus appear for very low (cw) excitations \cite{Chen2018,Ye2018,Li2018,Barbone2018}, are consistent with our findings about efficient bright-dark exciton scattering. 

Third, we discuss \textit{environment charge effects.} Standard TMD samples tend to contain residual charge carriers, which are mostly electrons resulting from chalcogenide vacancies. Some of them are trapped at localization centers and are likely activated via photoinjection. The moving charge carriers would serve as efficient scattering centers for excitons, which suffer from Auger-type annihilation. We attribute the observed slow dynamics with a time constant of $\sim250$~$\mu$s to the charge fluctuation process. 
Note that recent observations of near-unity quantum yields in field-effect devices also suggest the strong impact of background charge densities on exciton recombination \cite{Lien_Science19}. 
Nevertheless, we cannot exclude other environment mechanisms as possible origins of the observed slow dynamics. Further studies using a charge-tunable and hBN-encapsulated device would help to determine the influence of the environment on radiative dynamics. 

Here, we considered three representative long-lived states, i.e., large momentum excitons, spin triplets, and photo-activated free carriers, as a source of the measured slow relaxation processes. These states are non-emissive and `\textit{dark}'. Thus, their lifetimes are much longer than that of emissive `\textit{bright}' excitons. Hence, under stationary excitation, these long-lived states are populated more densely than bright excitons, and they can serve as efficient scattering centers for (bright) excitons, which are then annihilated nonradiatively through Auger-type energy transfer. We successfully quantified the lifetimes of the long-lived states by analyzing the process of luminescence recovery after stopping photo-excitation. Note that these long-lived states have physically independent origins. Thus, the experimentally measured curves simply follow the linear superposition of different curves, each of which is described by a single decay process. In our experiment, we observed luminescence signals that decay (or rise) with very different timescales (see, e.g., two decaying components in Fig.~\ref{fig_cwIntensity}), thus we can safely and precisely quantify 4 and 250~$\mu$s as their lifetimes. %


\section{Conclusion}
In conclusion, we studied exciton recombination dynamics in WS$_2$ monolayers capped with hBN over broad time scales from picoseconds to milliseconds and confirmed that luminescence efficiency was governed by Auger scattering between excitons and long-lived dark states. Newly designed pump-and-probe techniques enabled us to determine that the lifetimes of the optically inaccessible states were 4 and 250~$\mu$s. We attributed the state with the 4-$\mu$s lifetime to spin-forbidden dark excitons, and the state with the 250-$\mu$s lifetime to photo-activated environment charge carriers. Hence, we found that Auger annihilation occurs even for modest injections, which could easily be achieved in standard light emitting devices. A potential way of suppressing nonradiative exciton recombination is to realize a clean environment that is free from residual charge carriers and localization centers by adopting hBN-encapsulated charge-tunable structures. 



%
%

%

\begin{acknowledgments}
We acknowledge the support of CREST, Japan Science and Technology Agency (JST) under grant number JPMJCR15F3 and Japan Society for the Promotion of Science (JSPS) KAKENHI under grant numbers JP19H01820, JP20H00127, and JP20H00354. 
\end{acknowledgments}

\begin{figure}
\includegraphics[width=6cm]{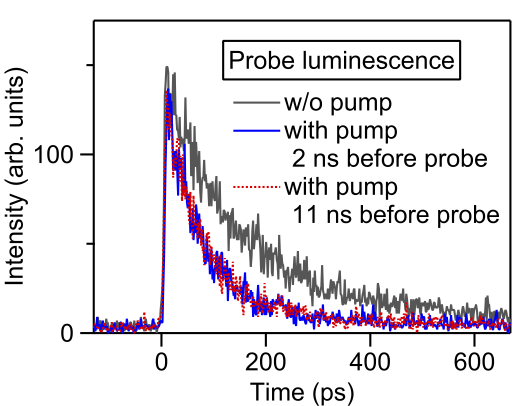}
\caption{\label{fig_2pulseLongDelay}%
Pump-and-probe response of luminescence decay signals. The gray line is a signal induced by probe pulses (50~nW) alone. The blue and red lines, respectively, indicate signals with pump pulses (200~nW) that arrived 2 and 11~ns prior to the probe pulses.%
}
\end{figure}

\appendix*\section{Effect of pump-and-probe delay time on luminescence decay signals}
In Sec.~\ref{subsec_PP}, we analyzed the luminescence decay signals excited with pump and probe pulses, which were separated by 1.44~ns (Fig.~\ref{fig_2pulse}). Here, we vary the pulse separation and observe its effect on the probe luminescence decay curve. Figure~\ref{fig_2pulseLongDelay} shows the luminescence decay signals induced by probe pulses. The gray line is a signal excited with probe pulses alone (without pump pulses). The blue curve is a signal where pump pulses are injected 2~ns prior to probe pulses. It reveals the acceleration in the luminescence decay, as also described in Sec.~\ref{subsec_PP}. The red line is a signal with pump pulses injected 11~ns prior to the probe pulses. The red and blue lines overlap perfectly, and there is no significant difference between them. Hence, we can conclude that the long-lived states, which cause Auger scattering, have lifetimes much longer than 11~ns. Thus, the observed long-lived states are in a stationary regime under 76~MHz excitation. To study such slow dynamics, we need to introduce an alternative excitation scheme that uses a fast modulated quasi-cw source, as described in Sec.~\ref{sec_qcwStudy}. 


\bibliography{darkstateWS2.bib}

\end{document}